\documentclass[onecollarge,runningheads]{svjour2}
\smartqed  
\usepackage{graphicx}

\journalname{Journal of Biological Physics}

\begin{document}

\title{Infection spreading in a population with evolving contacts}

\titlerunning{Infection spreading in a population with evolving contacts}

\author{Dami\'an H. Zanette   \and
        Sebasti\'an Risau Gusm\'an
}

\institute{D. H. Zanette and S. Risau Gusm\'an at \at Consejo
Nacional de Investigaciones Cient\'{\i}ficas y  T\'ecnicas, Centro
At\'omico Bariloche and Instituto Balseiro, 8400 San Carlos
de Bariloche, R\'{\i}o Negro, Argentina \\
Tel.: +54-2944-445173\\
Fax: +54-2944-445299\\
\email{zanette@cab.cnea.gov.ar, srisau@cab.cnea.gov.ar} }

\date{Received: date / Accepted: date}

\maketitle

\begin{abstract}
We study the spreading of an infection within an SIS epidemiological
model on a network. Susceptible agents are given the opportunity of
breaking their links with infected agents. Broken links are either
permanently removed or reconnected with the rest of the population.
Thus, the network coevolves with the population as the infection
progresses. We show that a moderate reconnection frequency is enough
to completely suppress the infection. A partial, rather weak
isolation of infected agents suffices to eliminate the endemic
state. 
\end{abstract}

\section{Introduction}
\label{intro}

Outbursts of epidemics in human populations trigger individual and
collective reactions that can substantially alter the social
structure. As a consequence of risk perception, non-infected
individuals may start avoiding contact with their infected equals,
even when their previous relationship was fluent. The whole society
could collectively decide to isolate its infected members until
danger is overcome. More altruistic non-infected individuals may be
tolerant of the contact with infected individuals but, in turn, the
latter may discontinue the relationship to impede contagion. The
escape from crowded cities during the Black Death in the late Middle
Ages, documented in Giovanni Boccaccio's Decameron, and the closing
of schools, churches, and theaters during influenza epidemics in the
early twentieth century, constitute dramatic historical instances of
such behaviours \cite{sigloxx,sigloxx1}. Quarantine protocols, and
preventive isolation during leprosy or tuberculosis treatment, are
present-day examples \cite{quar}. In any case, these changes in the
pattern of social contacts help to limit and control the incidence
of the infection.

In this paper, we explore the effects of an evolving pattern of
contacts on the dynamics of infection spreading, in the framework of
a simple epidemiological model. We consider a population of agents
whose pattern of contacts is represented by a network. If a link of
the network joins two agents, contagion is possible when one of them
is infected and his neighbour is not. To account for the social
processes addressed in the previous paragraph, we admit that the
contact network is not a static structure, but evolves in response
to the epidemiological state of the population.

Agent-based models whose interaction patterns are represented by
networks have received increasing attention during the last years,
in the analysis of emergent collective behaviour in complex systems
\cite{growth}. Frequently, the evolution of the interaction network
and the dynamics of individual agents occur over different time
scales. In learning processes, for instance, connections change
adaptively over scales that are large as compared with the internal
dynamics of agents \cite{learning}. At the opposite limit, in models
of network growth, the pattern evolves in the absence of any
dynamics related to the agents  \cite{growth}. When, on the other
hand, the dynamical time scales of a population of agents and its
interaction network are comparable, we can speak about their {\em
coevolution} \cite{zim,stau,gil1,gil2,holme}. In this context, the
model considered in the present study can be regarded as an
illustration of the coevolution of agents and networks, inspired in
the dynamics of infection spreading.

Our model is based on an SIS epidemiological process where, at a
given time, each agent can be susceptible (S) or infected (I). In
the standard SIS process, each I-agent spontaneously recovers and
becomes susceptible at a fixed rate, say, with probability $\gamma$
per unit time. An S-agent, in turn, becomes infected by contagion
from his infected neighbours. If the contagion probability per unit
time and per infected neighbour is $\rho$, an S-agent with $k_{\rm
I}$ infected neighbours becomes itself infected with probability
\begin{equation} \label{pI}
p_{\rm I} dt = 1-(1-\rho \ dt)^{k_{\rm I}} \to k_{\rm I} \rho \ dt
\end{equation}
during the interval $dt$.  Within a mean-field description, if the
average number of (both S and I) neighbours per agent is $k$ and the
fraction of I-agents is $n_{\rm I}$, we have $ k_{\rm I}= kn_{\rm
I}$. The mean-field evolution equation for $n_{\rm I}$  thus reads
\begin{equation} \label{sis0}
\dot n_{\rm I}  = -\gamma n_{\rm I} +  \lambda n_{\rm I} n_{\rm S},
\end{equation}
where $n_{\rm S}=1-n_{\rm I}$ is the fraction of S-agents, and
$\lambda =k \rho$. In this description, for asymptotically long
times, $n_{\rm I}$ vanishes if $ \lambda \le \gamma$. Therefore, the
infection is suppressed as time elapses. If, on the other hand, $
\lambda > \gamma$, the fraction of infected agents approaches a
finite value $n_{\rm I}^*=1-\gamma/ \lambda>0$, and the infection is
endemic. The transition between these two regimes occurs through a
transcritical bifurcation at $ \lambda = \gamma$.

In the following, we complement the standard SIS model with the
possibility that the network of contacts changes in response to the
infection spreading. Specifically, links between susceptible and
infected agents can be broken, and either removed or reconnected to
other agents. As expected, we find that this mechanism decreases the
infection level, and can eventually suppress the endemic state. With
respect to the standard model, however, infection suppression for
high rates of contact change occurs through a tangent bifurcation,
which in turn gives rise to a bistability regime. In this regime,
the infection persists or dies out depending on the initial fraction
of infected agents. More unexpectedly, infection suppression does
not require a drastic overall change in the network structure, but
is reached with a moderate unbalance between the mean number of
neighbours of susceptible and infected agents. As we discuss in the
final section, these features are robust under several variations of
the dynamical rules.

\section{Evolution of the network of contacts: Link removal}
\label{sec:rem}

We address first the case where before the interaction between an
S-agent and an I-agent joined by a network link effectively takes
place, so that contagion becomes possible, they may decide to
definitively delete that link, thus avoiding any further contact. In
this situation, as far as the number of I-agents does not vanish,
the network of contacts keeps loosing its links. According to our
discussion of the standard SIS model (\ref{sis0}), however, we
expect that when the number of neighbours per agent has decreased
sufficiently, the infection dies out. Once no I-agent remains in the
population, removal of links stops and the systems reaches a static,
fully healthy state. To quantify the mechanism of link removal, we
assume that each link between an S-agent and and I-agent is deleted
with probability $q$ per time unit.

Our formulation of the dynamics of the present model is analogous to
the mean-field approach used to derive Eq. (\ref{sis0}). We assume
that the population consists of $N$ agents, and call $N_{\rm I}$ and
$N_{\rm S}$ the number of I and S-agents, respectively, so that
$N=N_{\rm I}+N_{\rm S}$. Additionally, we must introduce new
variables to describe the structural state of the network and its
relation with the epidemiological state of the population.
Therefore, we consider the quantities $M_{\rm II}$, $M_{\rm IS}$,
and $M_{\rm SS}$, respectively, the number of network links joining
two infected agents (II), an infected agent and a susceptible agent
(IS), and two susceptible agents (SS). The total number of links is
$M=M_{\rm II}+M_{\rm IS}+M_{\rm SS}$. Due to removal of IS-links,
the average change in $M_{\rm IS}$ per time unit is
\begin{equation}
\dot M_{\rm IS} = - q M_{\rm IS}. \ \ \ \ \ \ \ \ \mbox{(link
removal)}
\end{equation}

The processes of recovery and infection are the same as discussed
for the standard SIS model (\ref{sis0}) in the Introduction.
Recovery of an I-agent occurs with probability $\gamma$ per unit
time and, for each S-agent, the infection probability per unit time
and per infected neighbour is $\rho$. In a recovery event, when an
I-agent becomes susceptible, there is not only a decay in the number
of I-agents, but also a change in $M_{\rm II}$, $M_{\rm IS}$, and
$M_{\rm SS}$. In fact, the links joining the recovered agent with I
and S-agents pass, respectively, from the II-type to the IS-type,
and from the IS-type to the SS-type. The number of links of each
type associated to a given agent is calculated using mean-field-like
averages. For instance, the number of II-links associated to an
I-agent is estimated as $2M_{\rm II} / N_{\rm I}$. Similarly, the
number of IS-links associated to an I-agent is $M_{\rm IS} / N_{\rm
I}$. Using this kind of arguments, we obtain, for each variable, the
average change per time unit due to recovery:
\begin{equation} \label{recov}
 \left.
\begin{array}{rl}
\dot  N_{\rm I} &=- \dot N_{\rm S}=-\gamma N_{\rm I} , \\
\dot  M_{\rm II}&=-2\gamma M_{\rm II}  ,\\
\dot  M_{\rm IS}&=2\gamma M_{\rm II}-\gamma M_{\rm IS}, \\
\dot  M_{\rm SS}&=\gamma M_{\rm IS} .
 \end{array} \right. \ \ \ \ \ \mbox{(recovery)}
\end{equation}
Note that $\dot  M_{\rm II}+\dot  M_{\rm IS}+\dot M_{\rm SS}=0$,
because recovery events do not change the number of network links.

To calculate the contribution of infection events in the mean-field
approximation, we must evaluate the average number of infected
neighbours of a susceptible agent. For a randomly chosen S-agent,
this number is given by the ratio $M_{\rm IS}/N_{\rm S}$. However,
it should be taken into account that, to become infected by
contagion, a susceptible agent must have at least one infected
neighbour. This would restrict the calculation of the number of
infected neighbours to those S-agents with at least one IS-link. For
the sake of simplicity, we shall still estimate the average number
of infected neighbours per S-agent as the above ratio, with the
proviso that the approximation is valid when the overall number of
I-agents is not too small, so that contagion is in principle
possible for all S-agents.   The change per time unit due to
infection for each variable turns out to be
\begin{equation} \label{infec}
 \left.
\begin{array}{rl}
\dot  N_{\rm I} &=- \dot N_{\rm S}= \rho M_{\rm IS} , \\
\dot  M_{\rm II}&= \rho M_{\rm IS}^2 /N_{\rm S},\\
\dot  M_{\rm IS}&=\rho (2M_{\rm SS}-M_{\rm IS}) M_{\rm
IS}/N_{\rm S}, \\
\dot  M_{\rm SS}&=-2 \rho M_{\rm SS} M_{\rm IS}/N_{\rm S} .
 \end{array} \right. \ \ \ \ \ \mbox{(infection)}
\end{equation}
Again, $\dot  M_{\rm II}+\dot  M_{\rm IS}+\dot M_{\rm SS}=0$.

To obtain differential equations of the type of Eq. (\ref{sis0}), it
is convenient to define the fractions $n_i = N_i/N$ and
$m_{ij}=M_{ij}/M$, with $\{ i,j \} \equiv \{ {\rm I}, {\rm S}\}$. In
calculating the variation of $m_{ij}$ per time unit, we must take
into account that also the total number of links $M$ varies with
time:
\begin{equation}
\dot m_{ij} = \frac{\dot M_{ij}}{M} -\frac{M_{ij}}{M^2} \dot M.
\end{equation}
Since the total number of links changes by the removal of IS-links
only, we have $\dot M = -q M_{\rm IS}$. On the other hand, the total
number of agents $N$ remains constant. Also, by definition, we have
$n_{\rm I}+n_{\rm S}=1$ and $m_{\rm II}+m_{\rm IS}+m_{\rm SS}=1$, so
that we can limit ourselves to study the evolution of $n_{\rm I}$,
$m_{\rm II}$, and $m_{\rm IS}$.

The evolution equations resulting from the above considerations are
\begin{equation} \label{eqsrem}
\begin{array}{rl}
n_{\rm I}'&=-n_{\rm I}+ \tilde \lambda  m_{\rm IS}, \\
m_{\rm II}'&= -2m_{\rm II}+\tilde \lambda n_{\rm S}^{-1}m_{\rm
IS}^2+\tilde q m_{\rm II}m_{\rm IS}, \\
m_{\rm IS}'&=2m_{\rm II}-(1+\tilde q) m_{\rm IS}+\tilde \lambda
n_{\rm S}^{-1}m_{\rm IS}(2m_{\rm SS}-m_{\rm IS}) +\tilde q m_{\rm
IS}^2,
\end{array}
\end{equation}
where primes indicate differentiation with respect to the rescaled
time $t'=\gamma t$. We have also defined
\begin{equation}
\tilde q = q/\gamma, \ \ \ \ \ \tilde \lambda = k \rho/ 2\gamma,
\end{equation}
where $k=2M/N$ is the overall mean number of neighbours per agent.
Note that, since the number of links $M$ varies with time, the
coefficient $\tilde \lambda$ is itself time-dependent. Its evolution
is determined by the variation of $M$, and obeys
\begin{equation} \label{rhorem}
\tilde \lambda ' =-\tilde q m_{\rm IS} \tilde \lambda .
\end{equation}

The numerical solution of Eqs. (\ref{eqsrem}) and (\ref{rhorem})
confirms the expectation that the infection dies out as a
consequence of the sustained removal of links. The fraction of
infected agents asymptotically vanishes with time and, accordingly,
$m_{\rm II}$ and $m_{\rm IS}$ also tend to zero. Meanwhile, the
total number of links approaches a constant $M_R$. The number of
remaining links $M_R$ depends both on the initial number of links,
$M_0$, and on the initial fraction of infected agents. In fact, for
a given rate of link removal, the suppression of a higher initial
infection level is expected to take longer times and require more
removed links.

\begin{figure}
\centering \resizebox{.6\columnwidth}{!}{\includegraphics*{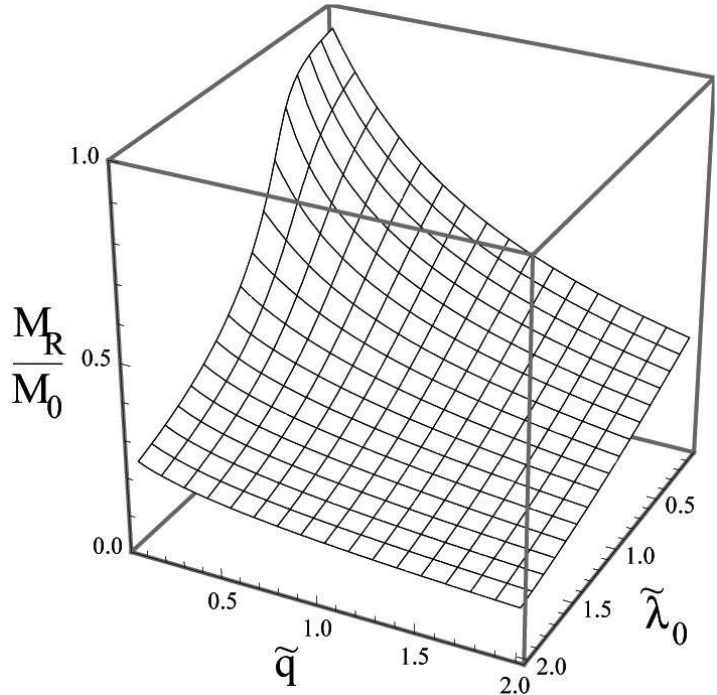}}
\caption{Number of remaining links, $M_R$, relative
to the initial number of links $M_0$, as a function of the
normalized removal probability $\tilde q$ and initial infectivity
$\tilde \lambda_0$, for an initial condition where all agents are
infected and the population is fully connected.} \label{frem}
\end{figure}

Figure \ref{frem} illustrates the dependence of the remaining
fraction of links, $M_R/M_0$, on the normalized rate of link removal
$\tilde q$ and on the infectivity. The infectivity is here
characterized by the initial value of $\tilde \lambda$, $\tilde
\lambda_0 = k_0 \rho/ 2 \gamma$, with $k_0=2M_0/N$. The results
correspond to an initial condition with a fully infected population,
$n_{\rm I}(0)=1$, and a fully connected network, $M_0=N(N-1)/2$, so
that $m_{\rm II}=1$ and $m_{\rm IS}=m_{\rm SS}=0$. As expected, the
fraction of remaining links decreases both with $\tilde \lambda_0$
and $\tilde q$. Higher infectivities require that a larger fraction
of contacts is deleted before the infection dies out, and a larger
removal probability contributes in the same direction.

From the viewpoint of the dynamics, the case considered so far
--where links keep being removed as long as a fraction of the
population remains infected-- is not especially interesting. In
particular, the transition between infection suppression and
persistence observed for $q=0$ when the infectivity  grows,
disappears as soon as the removal probability becomes positive. In a
real population, moreover, it is expected that discontinued contacts
are replaced, at least to some extent, by new social links, in such
a way that the structure of society is not too much deteriorated. In
the next section, we study a model where links between susceptible
and infected agents are not permanently removed, but rather
reconnected to other members of the population. Under these
conditions, both infection suppression and the endemic state are
possible, and can even coexist for a given set of parameters. The
dynamics is accordingly richer, and new critical phenomena
separating both regimes appear.

\section{Reconnection of links} \label{sec:rec}

We consider now that, before each susceptible-infected interaction
--possibly leading to contagion-- takes place, the susceptible agent
is given the opportunity of breaking the contact with his infected
neighbour and to reconnect the corresponding link to another agent,
randomly chosen from the rest of the population. Reconnection of
each IS-link occurs with probability $r$ per time unit. If the
susceptible agent is reconnected to another susceptible agent, the
link changes from the IS-type to the SS-type. Otherwise, no change
occurs.

A similar model was considered recently \cite{thg} where, however,
reconnection of S-agents always occurs towards other S-agents. In
this model, upon reconnection, IS-links always change to SS-links.
This variant is implicitly admitting that agents have information on
the (S or I) state of their equals before making contact, which
seems to be a rather artificial assumption. On the other hand, in
our model reconnection is done at random, which overcomes such
assumption but, at the same time, limits the efficiency of I-agent
isolation.

Only the variables $M_{\rm IS}$ and $M_{\rm SS}$ change due to
reconnection events. Since the probability of choosing an S-agent at
random is $n_{\rm S}$ we have, per unit time,
\begin{equation}
\dot M_{\rm IS}=-\dot M_{\rm SS}=-r n_{\rm S} M_{\rm IS}. \ \ \ \ \
\  \mbox{(link reconnection)}
\end{equation}
Changes due to recovery and infection are the same as in Eqs.
(\ref{recov}) and (\ref{infec}), respectively.

Obtaining the evolution equations for the fractions $n_{\rm I}$,
$m_{\rm II}$, and $m_{\rm IS}$, is now simpler than in the case of
link deletion, because the total number of links $M$ remains
constant. We find
\begin{equation} \label{eqsrec}
\begin{array}{rl}
n_{\rm I}'&=-n_{\rm I}+ \tilde \lambda  m_{\rm IS}, \\
m_{\rm II}'&= -2m_{\rm II}+\tilde \lambda n_{\rm S}^{-1}m_{\rm IS}^2, \\
m_{\rm IS}'&=2m_{\rm II}-(1+\tilde r n_{\rm S}) m_{\rm IS}+\tilde
\lambda n_{\rm S}^{-1}m_{\rm IS}(2m_{\rm SS}-m_{\rm IS}).
\end{array}
\end{equation}
Here, again, primes indicate differentiation with respect to the
rescaled time $t'=\gamma t$. Moreover,
\begin{equation}
\tilde r=r/ \gamma,  \ \ \ \ \ \tilde \lambda = k \rho /2 \gamma,
\end{equation}
with $k=2M/N$ the average number of neighbours per agent. Since,
now, $M$ does not vary with time, both $k$ and $\tilde \lambda$ are
constants. Note that the normalized reconnection probability $\tilde
r$ and infectivity $\tilde \lambda$ are the only parameters of Eqs.
(\ref{eqsrec}). The normalized infectivity incorporates the only
network-specific feature, namely, the mean connectivity $k$.

\subsection{Infection level at equilibrium}

We focus the attention on the equilibrium solutions of Eqs.
(\ref{eqsrec}), which are the candidates to represent the infection
level and network structure at asymptotically long times. First, we
consider the stationary values of the fraction of infected agents.
The analysis is restricted to the case of $\tilde \lambda>1/2$
which, in the absence of reconnection events ($r=0$), corresponds to
an endemic infection, $n_{\rm I} > 0$ for $t \to \infty$ (cf. the
discussion of the standard SIS model in the Introduction).

At the fixed points of Eqs. (\ref{eqsrec}), the equilibrium
fractions of links $m_{\rm II}^*$ and $m_{\rm IS}^*$ are related to
the equilibrium fraction of infected agents $n^*_{\rm I}$ as
\begin{equation} \label{ms}
m_{\rm II}^*=\frac{n^{*2}_{\rm I}}{2\tilde \lambda (1-n^*_{\rm I})},
\ \ \ \ \ \ \ \ m_{\rm IS}^*= \frac{n^*_{\rm I}}{\tilde \lambda}.
\end{equation}
In turn, $n^*_{\rm I}$ satisfies
\begin{equation} \label{ni0}
0= n_{\rm I}^* \left[ 2\tilde \lambda -1-\tilde r +(3\tilde
r-2\tilde \lambda) n_{\rm I}^* -3\tilde r n_{\rm I}^{*2} +\tilde r
n_{\rm I}^{*3} \right].
\end{equation}
This polynomial equation has four solutions. One of them tends to
infinity for $\tilde r \to 0$, and remains real and larger than one
for any positive $\tilde r$. Since meaningful solutions to our
problem must verify $n_{\rm I}^*\le 1$, we disregard this solution
from now on.

The trivial solution $n_{\rm I}^{(0)}=0$ exists for any value of the
normalized infectivity $\tilde \lambda$ and of the normalized
reconnection probability $\tilde r$. For a given $\tilde \lambda$,
its stability depends on $\tilde r$. As discussed in more detail
below, $n_{\rm I}^{(0)}=0$ is unstable for small $\tilde r$ and
becomes stable as $\tilde r$ grows. The other two solutions read
\begin{equation} \label{n12}
n_{\rm I}^{(1,2)} = 1-\sqrt{\frac{2\tilde \lambda}{3\tilde r}}
\left[ \cos \frac{\alpha}{3} \mp \sqrt{3} \sin \frac{\alpha}{3}
\right] ,
\end{equation}
with
\begin{equation}
\alpha = \arctan \sqrt{\frac{32 \tilde \lambda^3}{27 \tilde r}-1}
\end{equation}
($0\le \alpha \le \pi/2$). These two solutions are real for $32
\tilde \lambda^3 \ge 27 \tilde r$. Otherwise, they are complex
conjugate numbers. The solution $n_{\rm I}^{(1)}$, with the minus
sign in the right-hand side of Eq. (\ref{n12}), approaches
$1-(2\tilde \lambda)^{-1}$ for $\tilde r \to 0$. Thus, it represents
the expected fraction of infected agents in the absence of
reconnection. When it is real, it satisfies $n_{\rm I}^{(1)}<1$, and
it is stable as long as it remains positive. Consequently, along
with the trivial solution, $n_{\rm I}^{(1)}$ is another meaningful
equilibrium solution to our problem. Finally, $n_{\rm I}^{(2)}$ is
negative and stable for small $\tilde r$. Depending on  $\tilde
\lambda$, it can become positive as $\tilde r$ grows but, at the
same time, it becomes unstable. Therefore, it does not represent a
meaningful solution.

\begin{figure}
\centering
\resizebox{.8\columnwidth}{!}{\includegraphics*{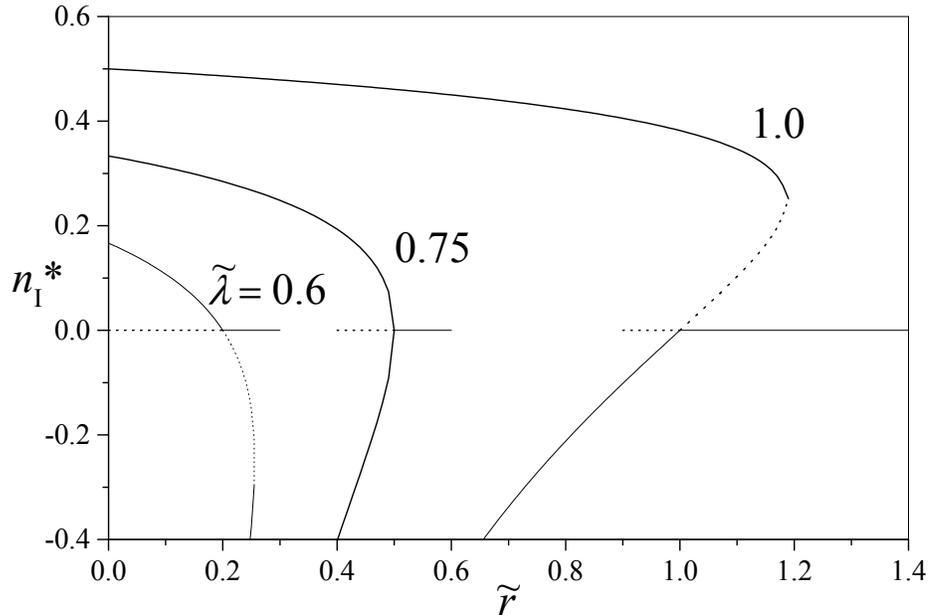}}
 \caption{Bifurcation diagram for the equilibrium fraction
of infected agents $n_{\rm I}^*$ as a function of the normalized
reconnection probability $\tilde r$, for three normalized
infectivities $\tilde \lambda$. Although only positive values of
$n_{\rm I}^*$ are meaningful, an interval in the negative domain is
also shown for completeness.  Full and dotted lines represent,
respectively, stable and unstable branches. For clarity, the
solution $n_{\rm I}^{(0)}=0$ is plotted in the vicinity of the
transcritical bifurcation only.} \label{fig1}
\end{figure}

Figure \ref{fig1} summarizes, in a bifurcation diagram, the
behaviour of $n_{\rm I}^{(0)}$, $n_{\rm I}^{(1)}$, and $n_{\rm
I}^{(2)}$ as functions of the normalized reconnection probability
$\tilde r$, for three representative values of the normalized
infectivity $\tilde \lambda$. In the three cases, we have $\tilde
\lambda>1/2$, so that --as discussed above-- a non-trivial
meaningful solution does exist. Full and dotted lines represent,
respectively, stable and unstable branches. For small infectivity
($\tilde \lambda = 0.6$), the stable solution $n_{\rm I}^{(1)}$
crosses $n_{\rm I}^{(0)}$ and becomes negative and unstable, while
$n_{\rm I}^{(0)}$ becomes stable. This transcritical bifurcation
takes place at $\tilde r = 2\tilde \lambda -1$. As $\tilde r$ grows
further,  $n_{\rm I}^{(1)}$ and the negative stable solution $n_{\rm
I}^{(2)}$ approach each other, and collide when $n_{\rm I}^{(1)} =
n_{\rm I}^{(2)}= 1-3/4\tilde \lambda$. Beyond this tangent
bifurcation, which takes place at $\tilde r = 32 \tilde \lambda^3 /
27$, the two solutions are complex numbers.

The situation is different for larger infection probabilities, as
illustrated by Fig. \ref{fig1} for $\tilde \lambda=1$. Now, for
$\tilde r=0$,  $n_{\rm I}^{(1)}$ is large and, as $\tilde r$ grows,
it is the stable negative solution $n_{\rm I}^{(2)}$ which first
reaches $n_{\rm I}^{(0)}$. At the transcritical bifurcation at
$\tilde r = 2\tilde \lambda -1$, $n_{\rm I}^{(2)}$ becomes positive
and unstable, and $n_{\rm I}^{(0)}$ becomes stable. The tangent
bifurcation where $n_{\rm I}^{(1)}$ and $n_{\rm I}^{(2)}$ collide
and become complex, at $\tilde r = 32 \tilde \lambda^3 / 27$, takes
now place when these two solutions are positive. As a consequence,
there is an interval of normalized reconnection probabilities,
between the two bifurcations, where the system is bistable: both
$n_{\rm I}^{(0)}$ and $n_{\rm I}^{(1)}$ are stable meaningful
solutions to the problem. The asymptotic state is selected by the
initial condition for $n_{\rm I}$.

The regimes of small and large infectivity are separated by the
critical value $\tilde \lambda =3/4=0.75$, also shown in Fig.
\ref{fig1}. At this critical point, the transcritical bifurcation
and the tangent bifurcation collapse into a  pitchfork bifurcation
at $\tilde r = 1/2$. Here, the three equilibria collide
simultaneously, and $n_{\rm I}^{(0)}$ becomes stable, while the
other two solutions become complex.

\begin{figure}
\centering
\resizebox{.8\columnwidth}{!}{\includegraphics*{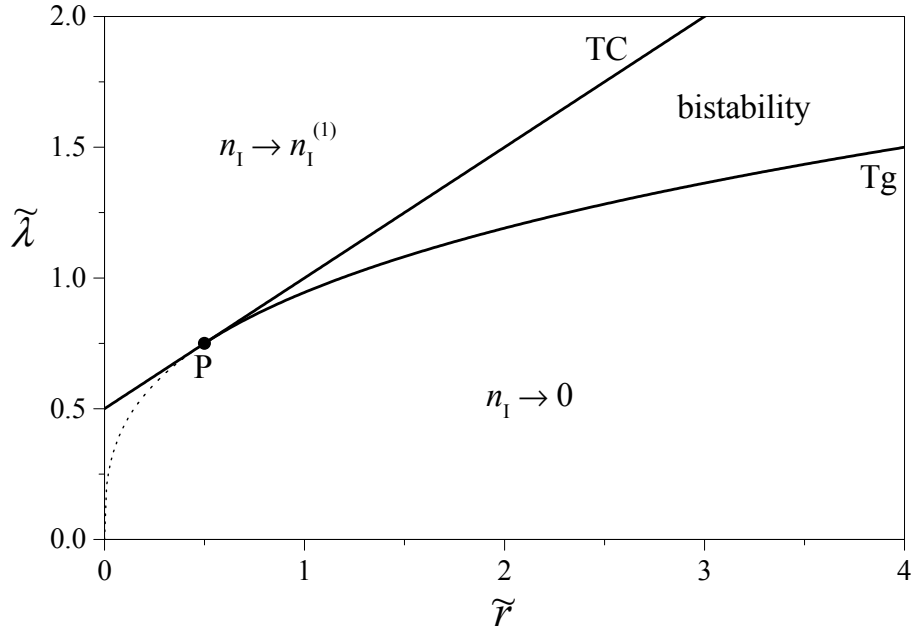}}
\caption{Phase diagram in the $(\tilde r, \tilde \lambda)$-plane,
showing the regions of infection suppression ($n_{\rm I} \to 0$) and
persistence ($n_{\rm I} \to n_{\rm I}^{(1)}$), and the intermediate
bistability zone. Their boundaries are given by the transcritical
(TC) and the tangent (Tg) bifurcation lines, which collapse into a
pitchfork bifurcation (P) at $(1/2,3/4)$. The dotted line is the
continuation of the tangent bifurcation line in the zone where
$n_{\rm I}^{(1)}$ is negative.  } \label{fig2}
\end{figure}

A phase diagram of our system over the parameter plane ($\tilde
r,\tilde \lambda$) is shown in Fig. \ref{fig2}. The zones of endemic
infection, where the fraction of infected agents at asymptotically
long times is positive ($n_{\rm I}\to n_{\rm I}^{(1)}$), and of
infection suppression ($n_{\rm I}\to 0$) are separated, for large
$\tilde \lambda$ and $\tilde r$, by the bistability region, where
the two asymptotic behaviours can be obtained, depending on the
initial condition. The three zones are limited by the lines of the
transcritical bifurcation [$\tilde \lambda = (1+\tilde r)/2$, TC],
where $n_{\rm I}^{(0)}=0$ changes its stability, and of the tangent
bifurcation [$\tilde \lambda = (27 \tilde r/32)^{1/3}$, Tg] where
$n_{\rm I}^{(1)}$ and $n_{\rm I}^{(2)}$ collide and become complex.
These two lines are tangent to each other at the ``triple point''
$(1/2,3/4)$, where the bistability region disappears, and the system
undergoes a pitchfork bifurcation [P]. For smaller $\tilde \lambda$
and $\tilde r$ bistability is no more possible, and the zones of
infection persistence and suppression are separated by the
transcritical line. The tangent bifurcation takes now place at
negative values of $n_{\rm I}^{(1)}$ and $n_{\rm I}^{(2)}$ (dotted
line).

Let us summarize our results on the persistence or suppression of
the infection in terms of the non-normalized parameters. First, for
small infectivity, $\rho \le \gamma/k$, the infection is always
suppressed. In this situation, the infectivity is just too small to
sustain a finite infected population. For larger infectivities, on
the other hand, the infection can become established, depending on
the reconnection probability $r$. In the range $\gamma/k <\rho
<3\gamma/2k$, the infection is endemic if reconnections are
infrequent, $r<k\rho-\gamma$. Otherwise, for sufficiently frequent
reconnections, the infection dies out. The transition between the
two situations is continuous in the fraction of infected agents, and
occurs through a transcritical bifurcation. For even larger
infection probabilities, $\rho
>3\gamma/2k$, the regimes of persistence (low $r$) and
suppression (large $r$) are separated by a bistability zone, where
the infection persists or dies out depending on the initial fraction
of infected agents. The bistability zone is limited by the
transcritical bifurcation quoted above and a tangent bifurcation at
a reconnection probability  $r= 4 k^3 \rho^3 /27 \gamma^2 $. The
discontinuous nature of the tangent bifurcation implies that the
endemic state present in the bistability zone disappears abruptly at
the boundary, with a finite jump in the asymptotic fraction of
infected agents, from $n_{\rm I} = 1-\sqrt{k\rho /3r}>0$ to zero.

\subsection{Number of neighbours of infected and susceptible
agents}

The variables $m_{\rm II}$ and $m_{\rm IS}$ characterize how the
structure of the network is related to the state of the agents.
Reconnection events favor the growth of the number of SS-links at
the expense of IS-links. Thus, for $r>0$, S-agents should
asymptotically posses relatively large numbers of neighbours. The
equilibrium values  $m_{\rm II}^*$ and $m_{\rm IS}^*$ as functions
of the equilibrium fraction $n_{\rm I}^*$ of I-agents are given by
Eqs. (\ref{ms}). These equations show, as expected, that the
fraction of links connecting I-agents with any other agent is
proportional to the fraction of I-agents itself.

\begin{figure}
\centering
\resizebox{.8\columnwidth}{!}{\includegraphics*{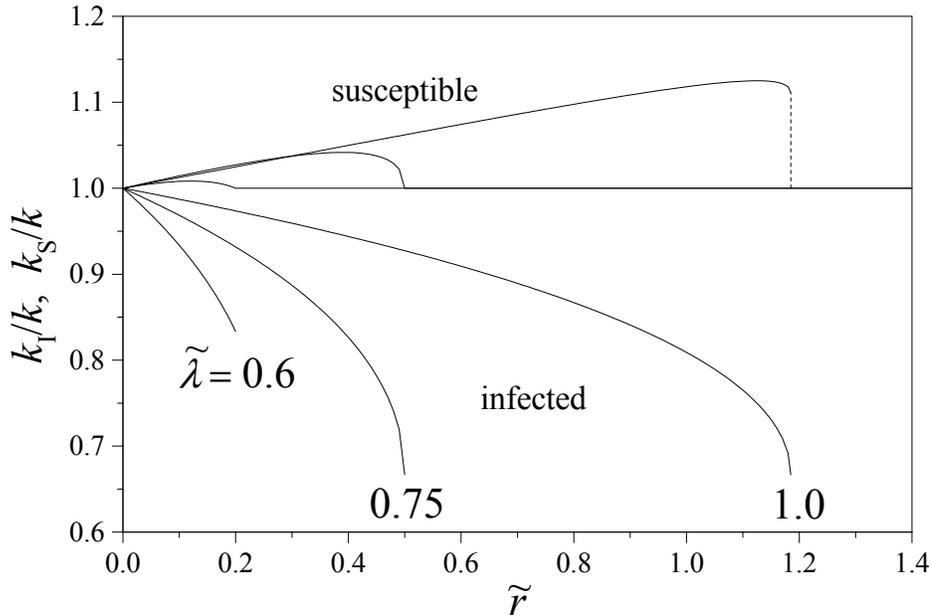}}
\caption{Connectivity of infected and susceptible agents, $k_{\rm
I}$ and $k_{\rm S}$, relative to the overall connectivity $k=2M/N$,
for three values of the normalized infectivity $\tilde \lambda$, as
functions of the normalized reconnection probability $\tilde r$.
Only the values corresponding to meaningful stable solutions for the
fraction of infected agents at plotted. The connectivity of infected
agents is not plotted beyond the threshold of infection suppression.
The vertical dashed line represents the finite jump in $k_{\rm S}$
at the tangent bifurcation where the solution $n_{\rm I}^{(1)}$
disappears.} \label{fig3}
\end{figure}

In order to introduce quantities that define the connectivity of
I-agents and S-agents independently of their respective fractions,
we consider the average number of neighbours per agent of each type.
For I-agents, for instance, the average numbers of infected and
susceptible neighbours are $2M_{\rm II}/N_{\rm I}$ and $M_{\rm
IS}/N_{\rm I}$, respectively. The average connectivity of I-agents,
$k_{\rm I}$, is the sum of these two quantities or, equivalently,
\begin{equation}\label{zi}
\frac{k_{\rm I}}{k} = \frac{1}{2\tilde \lambda (1-n_{\rm I}^*)},
\end{equation}
which gives the ratio between $k_{\rm I}$ and the overall average
connectivity per agent, $k=2M/N$. With analogous arguments for
S-agents, their average connectivity reads
\begin{equation} \label{zs}
\frac{k_{\rm S}}{k} = \frac{1}{1-n_{\rm I}^*}-\frac{n_{\rm
I}^*}{2\tilde \lambda (1-n_{\rm I}^*)^2} .
\end{equation}
Due to the conservation of the total number of links, $k_{\rm I}$
and $k_{\rm S}$ are univocally related. This relation can be
obtained from Eqs. (\ref{zi}) and (\ref{zs}) by eliminating $n_I^*$,
which yields
\begin{equation}
k_{\rm S} = k_{\rm I}[1+2\tilde \lambda (1-k_{\rm I}/k)].
\end{equation}
In order to describe the correlation between the structure and the
state of the population it is however useful to analyze both $k_{\rm
I}$ and $k_{\rm S}$ as functions of the relevant parameters. Figure
\ref{fig3} illustrates the behaviour of $k_{\rm I}$ and $k_{\rm S}$,
as described in the following, for the infection probabilities
$\tilde \lambda$ already considered in Fig. \ref{fig1}.

For $n_{\rm I}^* = n_{\rm I}^{(1)}$, which stands for the stable
equilibrium solution for low reconnection probabilities, both
$k_{\rm I}$ and $k_{\rm S}$ approach $k$ as $\tilde r \to 0$. As
expected, in the absence of reconnection events, there is no
difference in the number of neighbours of infected and susceptible
agents. As $\tilde r$ grows from zero, we have $k_{\rm I}<k<k_{\rm
S}$. We thus verify that reconnection tends to increase the
connectivity of S-agents at the expense of I-agents.

The other solution relevant to the process, $n_{\rm I}^*=n_{\rm
I}^{(0)}=0$, corresponds to a purely susceptible population.
Accordingly, we find $k_{\rm S}=k$. Note also that Eq. (\ref{zi})
predicts $k_{\rm I} = k/2\tilde \lambda$, but this value is never
realized due to the total absence of I-agents in this state.

For $\tilde \lambda \le 3/4$, the fraction of I-agents decreases
monotonically with $\tilde r$ and vanishes continuously at the
transcritical bifurcation --or, for $\tilde \lambda = 3/4$, at the
pitchfork bifurcation. The connectivity of S-agents is $k_{\rm S}
=k$ both at $\tilde r=0$ and at the bifurcation. For intermediate
values of the reconnection probability $k_{\rm S}$ is larger than
$k$ and attains a maximum. This maximum, which at first glance may
seem surprising, can be easily explained. In fact, to sustain a
value of $k_{\rm S}$ larger than the overall average $k$, it is
necessary to have I-agents with a relatively low number of
neighbours. As the infection is progressively suppressed by
reconnection, the number of I-agents decreases and, accordingly,
their contribution to the average number of neighbours per agent
becomes less significant. At the bifurcation and beyond, S-agents
must account for the whole average, so that $k_{\rm S}$  returns to
its value for $\tilde r=0$, i.e. $k_{\rm S}=k$.

The connectivity of I-agents, in turn, is a monotonically decreasing
function of $\tilde r$, and reaches $k_{\rm I}=k/2\tilde \lambda <k$
at the bifurcation. This implies that, even at the threshold of
infection suppression, I-agents maintain a finite number of
neighbours within the population.

For $\tilde \lambda > 3/4$, again, the connectivity $k_{\rm S}$
associated with the solution $n_{\rm I}^{(1)}$ initially increases
with $\tilde r$, and attains a maximum. In the subsequent decay,
however, it does not reach $k_{\rm S}=k$. In fact, $n_{\rm I}^{(1)}$
disappears through a tangent bifurcation when it is still positive,
so that the jump in the infection level is discontinuous. At the
bifurcation, we find  $k_{\rm S}= 2k(2\tilde \lambda +3)/9>k$. The
connectivity of I-agents decreases with $\tilde r$ and, at the
bifurcation, its value is independent of $\tilde \lambda$: $k_{\rm
I}=2/3$.

From the viewpoint of the interplay of the epidemiological dynamics
and the structure of the underlying network, the most interesting
result of this analysis is the fact that the infection dies out even
when infected agents keep a substantial connectivity with the rest
of the population. In the cases illustrated in Fig. \ref{fig3}, for
instance, infected agents preserve more than $60$ \% of their
connections at the threshold where the infection level vanishes. In
other words, as we had already verified for link deletion,
reconnection needs not to completely isolate infected agents to
suppress the infection. A moderate, partial isolation of the
infected population is enough to asymptotically inhibit the endemic
state.

\section{Heuristic description of infection suppression by link reconnection}

What mechanisms are at work when the infection is suppressed by
reconnection, even when the connectivity of infected agents remains
fairly high? To advance an answer to this question, it helps to
consider a simpler dynamical system for the fraction of I-agents:
\begin{equation} \label{eff}
n_{\rm I}'=-n_{\rm I} +[2 \tilde \lambda - \tilde r (1-n_{\rm I})^2
]n_{\rm I} (1-n_{\rm I}).
\end{equation}
The right-hand side of this equation is just a rearrangement of that
of Eq. (\ref{ni0}). The equilibria of Eq. (\ref{eff}) are thus
identical to the equilibria for $n_{\rm I}$ in Eqs. (\ref{eqsrec}).
Moreover, their stability properties are also the same as in our
original system. It is important to understand, however, that
(\ref{eff}) and (\ref{eqsrec}) are not equivalent: they merely share
the same equilibrium behaviour in what regards the fraction of
I-agents.

We immediately see that Eq. (\ref{eff}) can be put in the form of
the standard mean-field equation (\ref{sis0}) for a SIS process if
we introduce the effective infection probability
\begin{equation}
\lambda_{\rm eff} = \gamma [2 \tilde \lambda - \tilde r (1-n_{\rm
I})^2 ]=  (1-r)k \rho -r (1-n_{\rm I})^2.
\end{equation}
In Eq. (\ref{sis0}) the threshold of infection suppression, where
the trivial equilibrium  changes its stability, is given by
$\lambda=\gamma$. Imposing this same condition to $\lambda_{\rm
eff}$, we find $\tilde r = 2\tilde \lambda-1$. But this is precisely
the suppression threshold in the system with reconnections.
Therefore, with respect to the stabilization of the trivial
equilibrium, the system (\ref{eqsrec}) is effectively equivalent to
the standard SIS model with infectivity $\lambda_{\rm eff}$. The
transcritical bifurcation of Eqs. (\ref{eqsrec}), where $n_{\rm
I}^{(0)}=0$ becomes stable, can be interpreted as a kind of
continuation for $r \neq 0$ of the transcritical bifurcation of the
SIS model without reconnection events.

\begin{figure}
\centering
\resizebox{.8\columnwidth}{!}{\includegraphics*{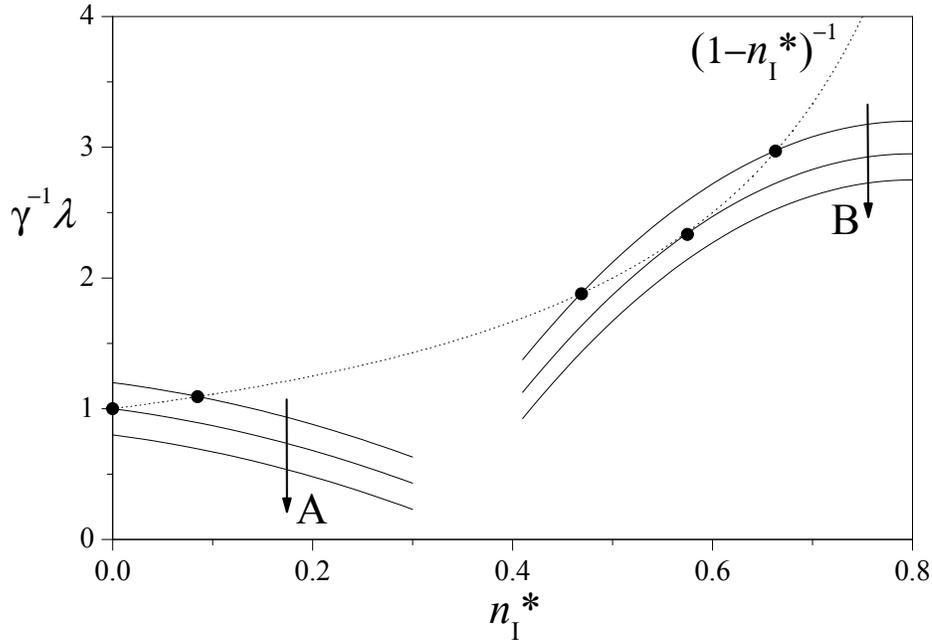}}
\caption{Graphical solution of Eq. (\ref{eff1}). The dotted curve
represents the right-hand side of the equation, and full curves are
possible graphs of the left-hand side. Dots stand at their
intersections. The arrows illustrate how the graphs may change upon
the variation of parameters, in the cases of a transcritical
bifurcation (A) and of a tangent bifurcation (B).} \label{fig4}
\end{figure}

The interpretation of the tangent bifurcation where the endemic
state disappears at a positive value of $n_{\rm I}^{(1)}$, for
$\tilde \lambda >3/4$, is less direct. It can however be argued that
the presence of such a  bifurcation, together with the transcritical
bifurcation which stabilizes the trivial equilibrium, constitutes
the most generic critical behaviour expected for an epidemiological
model like Eq. (\ref{sis0}) when the infection probability  depends
on the density of I-agents:
\begin{equation} \label{eff0}
n_{\rm I}'=-\gamma n_{\rm I} + \lambda (n_{\rm I}) n_{\rm I}
(1-n_{\rm I}).
\end{equation}
Besides the trivial equilibrium, this equation has fixed points at
the solutions of
\begin{equation} \label{eff1}
 \gamma^{-1}\lambda (n_{\rm I}^*) = (1-n_{\rm I}^*)^{-1}.
\end{equation}
Figure \ref{fig4} illustrates graphically two representative
situations. The dotted curve is the graph of the right-hand side of
Eq. (\ref{eff1}) as a function of $n_{\rm I}^*$. If the graph of
$\gamma^{-1} \lambda (n_{\rm I}^*)$ has a single intersection with
the dotted curve (A) and if, upon variation of parameters in the
infection probability, the graph varies as indicated by the arrow,
the intersection crosses $n_{\rm I}^*=0$ and a transcritical
bifurcation takes place. The standard SIS model, in which $\lambda$
is constant, is an example of this situation. More generally, the
graph of $\gamma^{-1} \lambda (n_{\rm I}^*)$ may have two (B) or
more intersections with the dotted curve. When the parameters
change, it is still possible than one of the intersections becomes
involved in a transcritical bifurcation crossing $n_{\rm I}^*=0$, as
in situation A. Now, however, it may well be the case that two
intersections approach each other, and eventually collapse and
disappear, as in B. In this case, Eq. (\ref{eff0}) undergoes a
tangent bifurcation, as found to happen in our system
(\ref{eqsrec}).

In summary, the above discussion shows that the suppression of the
endemic state as a result of reconnection events can be
heuristically understood in terms of the critical behaviour of a
standard SIS model with an effective infectivity, which depends on
both the reconnection probability and on the fraction of infected
agents. The transcritical bifurcation which stabilizes the state
where the infection is completely inhibited is interpreted as a
continuation of a similar transition in the absence of reconnection.
In turn, the tangent bifurcation --which, for large infectivities,
suppresses the infection as the reconnection probability grows-- is
a generic phenomenon in SIS models with density-dependent
infectivity \cite{tangent}.

\section{Conclusion}

In this paper, we have studied a model for an SIS epidemiological
process in a population of agents on a network, where contagion can
occur along the network links. The network coevolves with the
population as the infection progresses: as a response to risk
perception, susceptible agents can decide to break links with their
infected peers. In the first version of the model, broken links are
permanently removed from the network. For any positive probability
of link removal, the infection is found to asymptotically die out.
During the process, a fraction of network links is deleted, so that
the social structure is degraded in the long-time limit. As
expected, the fraction of remaining links decreases as the removal
probability and the infectivity grow.

In the second version, a susceptible agent who has broken a link
with an infected agent reconnects it to a randomly chosen member of
the remaining population. In this case, whether the infection
persists or is asymptotically suppressed depends on the reconnection
probability. Suppression of the endemic state does not require full
isolation of the infected population. On the contrary, it can be
achieved while each infected agent preserves a substantial part of
the links with the rest of the population.

Reconnection of network links introduces new dynamical features with
respect to the standard SIS process. In particular, for sufficiently
high reconnection probabilities, the continuous transition
associated with infection suppression is replaced by a discontinuous
tangent bifurcation, where the infected fraction of the population
drops abruptly as the relevant parameters are changed. The
appearance of this new critical phenomenon is accompanied by the
creation of a bistability regime, where the infection can either
persist or die out depending on the initial fraction of infected
agents.

These results are remarkably robust under variations of the
dynamical rules of the model. Here, for instance, we have assumed
that link reconnection occurs, at each time step, before infection
events, so that contagion can only take place from those infected
agents who have retained their links. If this ordering is altered,
the position of the bifurcation --and, consequently, the threshold
of infection suppression-- change, but the overall qualitative
picture of Fig. \ref{fig2} is not modified. The same holds if
reconnected links are kept, with a certain probability, by infected
agents, instead of always being susceptible agents which retain
broken contacts.

As shown in Section \ref{sec:rec}, the SIS model with link
reconnections can be  analytically solved in equilibrium. To a large
extent, the possibility of explicitly writing the stationary
solution for the density of infected agents and the expression for
critical lines in the parameter space is a direct consequence of the
approximation done in Section \ref{sec:rem}, just before Eqs.
(\ref{infec}), on the average number of infected neighbours of
susceptible agents. As stated there, the approximation is valid as
long as the number of infected agents remains high. On the other
hand, we have pushed the solution to describe also the regime where
the infection dies out. The question thus arises on whether our
analytical solution still gives a reasonable description of the
epidemiological dynamics. To advance an answer, we have performed
agent-based numerical simulations of the SIS process on the evolving
network. Results exhibit quantitative differences with the
analytical solution, especially in the prediction of the critical
points. Those are precisely the zones where the analytical
description is expected to fail, because at those points the
infection is suppressed and the approximation breaks down.
Qualitatively, however, the numerical and analytical results for the
critical behaviour are the same. Simulations confirm the
prolongation of the transcritical bifurcation for non-zero
reconnection probabilities, and the appearance of the tangent
bifurcation, along with the bistability regime, as reconnections
become more frequent.

Numerical simulations automatically incorporate further dynamical
elements which are not present in a mean-field analytical
description, such as the creation of correlations between the
relative positions of susceptible or infected agents and the effects
of heterogeneity in the distribution of network links. Extensive
numerical results, as well as an improved analytical approach able
to take into account correlations induced by the spatial structure
of the network \cite{network}, will be presented in a forthcoming
paper \cite{future}.





\end{document}